\documentclass{mpe_report}

\usepackage{psfig,graphicx,epsfig}
\usepackage{color}
\usepackage{amsmath,amssymb,epic,eepic,array}

\begin{document}

\pagenumbering{arabic}
\setcounter{page}{112}

\renewcommand{\FirstPageOfPaper }{112}\renewcommand{\LastPageOfPaper }{115}
\def \gtw{\>\hbox{\lower.25em\hbox{$\buildrel >\over\sim$}}\>}
\def \ltw{\>\hbox{\lower.25em\hbox{$\buildrel <\over\sim$}}\>}

\title{Radio Emission Physics in the Crab Pulsar}
\author{J.~A.~Eilek \and T.~H.~Hankins}  
\institute{New Mexico Tech, Socorro NM 87801, USA}
\maketitle

\begin{abstract}
Our high time resolution observations of individual giant pulses in the Crab
pulsar show that both the time and frequency signatures of the interpulse are
distinctly different from those of the main pulse.  Giant main pulses  can
occasionally be  resolved into short-lived, relatively narrow-band
nanoshots.  We believe these 
nanoshots are produced by soliton collapse in strong plasma
turbulence.  Giant interpulses are very different.  Their dynamic
spectrum contains narrow, microsecond-long emission bands.
We have detected these proportionately spaced bands from 4.5 to 10.5 GHz.  The
bands cannot easily be explained by any current theory of pulsar radio
emission; we speculate on possible new models.
\end{abstract}

\section{Introduction}
 What is the pulsar radio emission mechanism? Does the same mechanism
 always operate? Three types of models
have been proposed to explain the  radio emission:  coherent charge
bunches, plasma masers and strong plasma turbulence ({\it e.g.},
Hankins {\it et al.} 2003, ``HKWE'').  Because each model
makes different predictions for the time signature of the emission,
our group has 
carried out ultra-high time resolution observations in order to compare the
observed time signatures to those predicted by the models.

We have focused on the 
Crab nebula pulsar, because its occasional, very strong  giant pulses are
ideal targets for our observations.  
The dominant features of this star's mean
profile are a  main pulse (MP) and an interpulse (IP).  Although the
relative amplitudes and detailed profiles of these features change
with frequency, they can be 
identified from low radio frequencies ($\ltw 300$ MHz) up to the optical and
hard X-ray bands (Moffet \& Hankins 1996).  Some 
models suggest that the MP and IP come from low altitudes, above the 
star's two magnetic poles.  Other models suggest they come from higher
altitudes, possibly relativistic caustics (Dyks {\it et al.} 2004)  which
connect to the two poles.  In either case, the 
physical conditions in the emission region should be similar, and one
would expect the same radio emission mechanism to be active in the IP
and the MP.  We were surprised, therefore, to find  that the IP
and MP have very different properties.   It seems likely that they
differ in their emission mechanisms, their propagation within the
magnetosphere, or both.

\section{Giant main pulses:  strong plasma turbulence}

We initially studied the MP at nanosecond time resolution, because it
is usually brighter, and because giant pulses are more common
at the rotation phase of the MP (Cordes {\it et al.} 2004).  We found that most
giant main pulses (GMPs)   
consist of one to several ``microbursts'', each lasting a few
microseconds at 5 GHz (HKWE). We  recently extended our 
observations to higher frequencies, where 2 GHz of bandwidth is
available at Arecibo.  We found the temporal structure of GMPs is the
same at higher frequencies, although the microburst duration is
typically shorter than at 5 GHz.  Figure 1 shows a typical
example. The  dynamic spectrum of the  
microbursts turns out to be broadband, filling our entire observing
bandwidth.  An occasional MP, however, 
contains much shorter, relatively narrow-band, ``nanoshots'' (HKWE;  also 
Figures 2 and 3).    Most of the time the nanoshots overlap, which is
consistent with previous modelling of pulsar emission as
amplitude-modulated noise; but in sparse GMPs the nanoshots can
sometimes be  individually resolved.

\begin{figure}[htb]
\centerline{
\includegraphics[width=0.6\columnwidth,angle=-90]
     {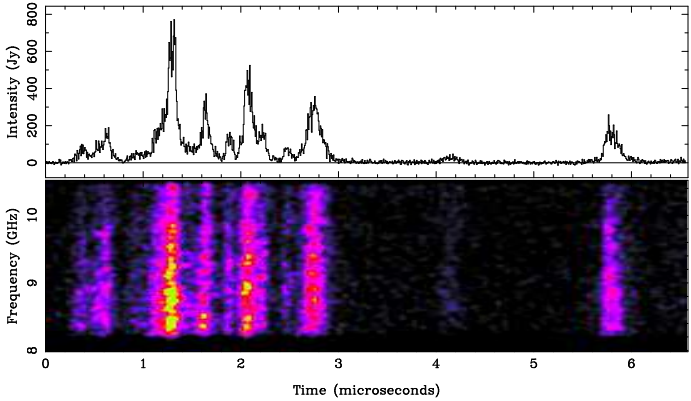}}
\caption{Total intensity and dynamic spectrum of a giant 
  main pulse,  observed at Arecibo and coherently dedispersed.
  As is typical   of most GMPs, it contains several broad-band
  microbursts. The   
  total intensity is plotted with 6.4-ns time resolution;  the dynamic
  spectrum has  25.6-ns time resolution and 19.5-MHz
 frequency resolution.}
\end{figure}

We used simple scaling arguments, and numerical simulations from
Weatherall (1998), to compare the nanoshots to predictions of the
three competing theoretical models of the radio emission mechanisms.  
The time signature of the nanoshots disagrees with predictions of the
maser and charge bunching models;  but both the time and frequency
signatures are consistent with Weatherall's  
numerical models of plasma emission by soliton collapse in strong
plasma turbulence.  His models predict nanoshot durations at frequency
$\nu$ to be $\nu \delta t  \sim O(10)$; an individual nanoshot is
relatively narrow-band, $\delta \nu /\nu \sim O(0.1)$. In HKWE we
suggested, based on the time signature of the nanoshots, that strong
plasma turbulence is the emission mechanism in GMPs.  The time and
spectral signatures of the nanoshots in our recent high-frequency work
are also consistent with these models.  We thus propose
that microbursts in giant main pulses are collections of nanoshots,
produced by strong plasma turbulence in the emission region.

\begin{figure}[htb]
\centerline{
\includegraphics[width=0.6\columnwidth,angle=-90]
     {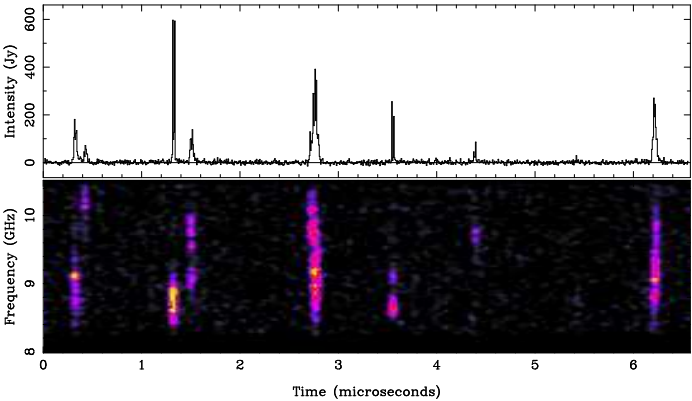}}
\caption{Another giant main pulse, processed and displayed the same as
  in    Figure 1, in which the microbursts are sparse and
  short-lived. At higher time 
  resolution some of  these bursts can be resolved into even shorter
  {\it  nanoshots}, as in Figure 3. Note the relatively narrow-band
  nature of individual nanoshots.   We
  infer that the longer-duration, broad-band  microbursts, such as those  in
  Figure 1,  are ``clouds'' of these nanoshots.  
}
\end{figure}

\begin{figure}[htb]
\centerline{
\includegraphics[width=0.6\columnwidth,angle=-90]
     {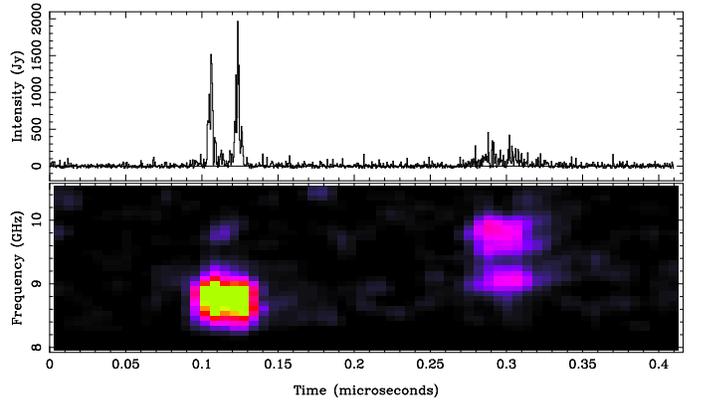}}
\caption{A short portion of the sparse main pulse shown in Figure 2, with
total intensity displayed at our maximum time resolution, 0.4 ns.  This reveals
the temporal signature of individual nanoshots.  We believe the nanoshots
are created by collapse of solitons in strong plasma turbulence.  The 
dynamic spectral resolution is 78 MHz and  6.4 ns.} 
\end{figure}

If our suggestion is correct, it has one important consequence.
Plasma flow in the radio emission region should be highly dynamic.
The plasma flow will be smooth only if the local charge density is
exactly  the Goldreich-Julian (GJ) value, so that the rotation-induced
electric field, $\mathbf E$, is fully shielded.  Because plasma
turbulent emission is centered on the comoving plasma 
frequency ($\nu_p \propto \sqrt{\gamma_b n}$, for number density $n$ and
bulk Lorentz factor $\gamma_b$), we can determine the local density in
the radio emission region ({\it cf.} also Kunzl {\it et al.} 1998).  We
find that low radio frequencies 
come from densities too low to match the GJ value anywhere in the
magnetosphere.  Because the emitting plasma feels an unshielded $\mathbf
E$ field, and feeds back on that field as its charge
density fluctuates, we expect unsteady plasma flow (and consequently
unsteady radio emission).  

\section{Giant interpulses:  emission bands}

In order to test our hypothesis that strong plasma turbulence governs
the emission physics in the Crab pulsar, we went to higher frequencies
to get a larger bandwidth and shorter time
resolution.  In addition to the MP, we observed giant pulses from the
IP, because at high frequencies giant pulses are more common at the
rotation phase of the IP. When we used 
the method described in HWKE to observe giant interpulses (GIPs) with
a broad  bandwidth, from 6-8 or 8-10 GHz, we were  astonished to find
that GIPs have very different 
properties from giant main pulses.  GIPs differ from GMPs in time
signature, polarization, dispersion and spectral properties. In this
paper we summarize our new results;  we will present
more details in a forthcoming paper (Hankins \& Eilek
2007). 
 
\subsection{Emission bands in the interpulse} 

The most
striking difference between the IP and the MP is found in the 
dynamic spectrum.   A giant IP contains  microsecond-long trains of 
{\em emission bands}, as illustrated in
 Figures 4 and 5.  The bands are grouped into regular ``sets'';
 2 or 3 band sets can usually be identified in a given IP.  Individual
band sets last a few $\mu$s.  In some pulses new band sets turn
on partway through the pulse, often coincident with a secondary burst
of total intensity. {\it Every} giant interpulse we recorded between 4.5
and 10.5 GHz, 
 during 20 observing days from 2004  to 2006, displays these emission bands.  
However, giant main pulses observed at the same
time and processed identically do not show the bands.  The bands are,
therefore,  not due to instrumental or interstellar effects, but are
intrinsic to the star.

\begin{figure}[htb]
\centerline{
\includegraphics[width=0.6\columnwidth,angle=-90]
   {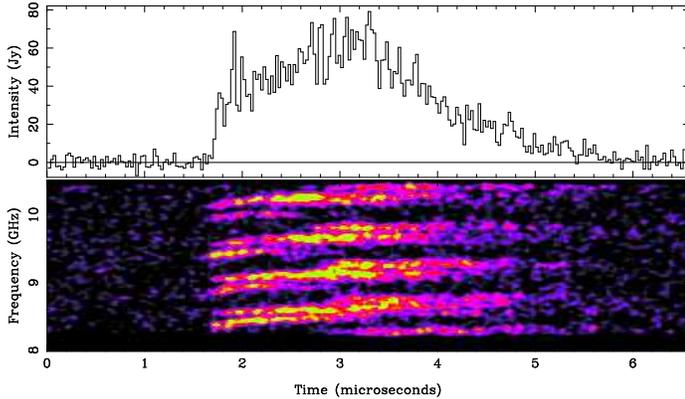}
}
\caption{A giant interpulse from the Crab pulsar, observed and processed 
in exactly the same way as the main pulses in Figures 1 through 3. 
Total intensity time resolution 25.6 ns;  dynamic spectral resolution
 19.5 MHz and 25.6 ns.  The banded 
frequency structure in the dynamic spectrum occurs in {\it every} IP we have
recorded above 5 GHz.  In this example, two band sets can be identified,
starting at 1 $\mu$s, each containing four bright bands;  a third band set
starts near 2 $\mu$s.  Note the apparent drift to higher frequencies
 for each band set. 
} 
\end{figure}

At first glance the bands appear to be uniformly spaced.  However, closer 
inspection of our data shows that the bands are {\it proportionally
  spaced}.  The spacing between two adjacent bands, at $\nu_1$
 and $\nu_2$, 
 depends on the mean frequency,  as $\Delta \nu/ \nu  = 2(\nu_2 -
 \nu_1)/( \nu_2 +  \nu_1) \simeq 0.06$.  Thus, two bands 
near 6 GHz are spaced 
by $\sim 360$ MHz;  two bands near 10 GHz are
spaced by $\sim 600$ MHz.  This proportional spacing is robust;  a set
of emission bands can drift in frequency (usually upwards, as in
Figures 3 and 4), but their frequency spacing stays constant.  All
bands in a particular set appear almost simultaneously, to within
$\sim 0.1 \mu$s;  they must all come from a region no larger than
$\sim 30$ m across.

We suspect the bands extend over at least a $5\!-\!6$ GHz range in a single
GIP, but do not occur below $\sim 4$ GHz. While we
have not been able to observe more than 2 GHz simultaneously, we have
seen no evidence that a given band set cuts off within our observable
bandwidth.  The characteristics of the bands (proportional spacing,
duration, onset relative to total intensity microbursts) are unchanged
from 5 to 10 GHz. In addition, the rotation phase of the
high-frequency IP is slightly
shifted relative to the low-frequency IP (Moffett \& Hankins
1996).  This phase offset suggests that the bands do not continue to
frequencies below $\sim 4$GHz.

\begin{figure}[htb]
\centerline{
\includegraphics[width=0.6\columnwidth,angle=-90]
   {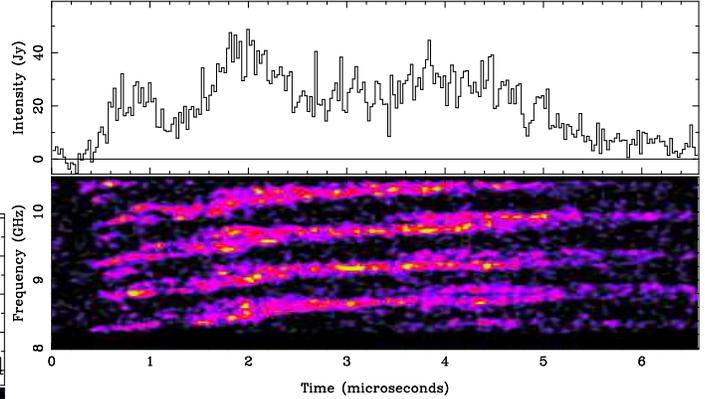}
}
\caption{Another giant interpulse, observed on   the same day as the
  IP in Figure 4 and processed and displayed in the same way. 
  Four emission band sets can be identified, including a set which
  coincides with the second microburst near $2\, \mu$s, and another
  which coincides with a third microburst near $4\, \mu$s.  This example
  also shows the characteristic band drift to higher frequencies.
} 
\end{figure}

\subsection{Possible causes of the emission bands}  

The dynamic spectrum of the giant interpulses does not match any of
 the  three types of emission models described above.  Because each of
 the models  predicts narrow-band emission at the plasma frequency,
 {\it none of them  can  explain the dynamic spectrum of 
the IP}.  A new approach is required here, which may ``push the
envelope'' of pulsar radio emission models.

 While we remain perplexed by the dramatic dynamic spectrum of the interpulse,
we are exploring possible models.  This exercise is made particularly 
difficult by the fact that the emission bands are not regularly
spaced. Because of this, models that initially seemed attractive must
be rejected. As an example, if the emission bands were uniformly
spaced they could be the spectral representation of a regular
emission pulse train.  Many authors have invoked
regularly spaced plasma structures (sparks or filaments), whose passage
across the line of sight could create such a pulse train.
Alternatively, strong plasma waves with a characteristic 
frequency will also create a regular emission pulse train.  The dynamic
spectrum of either of these models would contain emission bands at
constant spacing; the {\it proportional} spacing we observe disproves
both of these hypotheses. 

We have looked to solar physics for insight. We initially remembered
split bands in  the dynamic spectra of Type II solar flares, which are
thought to be plasma emission from low and high density regions
associated with  a shock propagating through the solar corona.  This
does not seem to be helpful for the Crab pulsar 
emission bands, because the radio-loud plasma would have to contain 10
or 15 different density stratifications, which seems 
unlikely.  However, ``zebra bands'' seen in Type IV solar flares
may be germane.  These are parallel, drifting, narrow emission bands
seen in the dynamic spectra of Type IV flares.  Band sets containing
from a few up to $\sim 30$ bands 
have been reported, with fractional spacing $\Delta \nu / \nu \sim .01
- .03$ ({\it e.g.}, Chernov {\it et al.} 2005).  While zebra bands have
not yet been satisfactorily explained, 
two classes of models have been proposed, invoking either resonant
plasma emission or geometrical effects.  Can similar models explain the
emission bands in the Crab pulsar?

{\it Resonant cyclotron emission.}  One possibility is plasma emission
at the cyclotron resonance, $\omega - k_{\parallel} v_{\parallel} - s
\Omega_o / \gamma = 0$ (where $\gamma$ is the particle Lorentz factor,
$\Omega_o = e B / m c$, and $s$ is the 
integer harmonic number).  Kazbegi {\it et al.} (1991) proposed that this
resonance operates at high altitudes in the pulsar magnetosphere, and
generates X mode waves which can escape the plasma directly.
Alternatively, ``double resonant'' cyclotron emission at the plasma
resonant frequency has been proposed for solar flares ({\it e.g.},
Winglee \& Dulk 1986).  In solar conditions, this resonance generates
O mode waves, which must mode convert in 
order to escape the plasma.  The emission frequency in these models is
determined by local conditions where the resonance is satisfied;   the
band separation  is $\Delta \nu \simeq \Omega_o / 2 \pi
\gamma$.  

Resonant emission  models face several challenges before they can be
considered successful.  The emission must  occur at high altitudes, in
order to bring the resonant (cyclotron) frequency  
down to the radio band.   Close to the light cylinder,
where $B \sim 3 \times 10^5$G, particle energies $\gamma \sim 10^3\!-\!
10^4$ are needed.
In addition, such models must be developed with  specific calculations
which address the fundamental 
plasma modes as well as their stability, under conditions likely to
exist at high altitudes in the pulsar's magnetosphere.  It is not
clear how the specific, proportional band spacing can be explained;
perhaps a local gradient in the magnetic field must be invoked.

{\it Geometrical models.}  Alternatively, the striking regularity of
the bands calls to mind a special geometry.  If some mechanism
splits the emission beam coherently,  so that it interferes with
itself, the bands could be interference fringes.  For instance, a
downwards beam which reflects off a high density region could return
and interfere with its upwards counterpart on the way back
up.  Simple geometry suggests that fringes occur if the two paths
differ in length by only $c / \Delta \nu \ltw 1$m. Another geometrical
possibility is that cavities form in the plasma and
trap some of the emitted radiation, imposing a discrete frequency
structure in the plasma ({\it   e.g.},  
LaBelle {\it et al.} 2003  for solar zebra bands).  The scales required
here are also small;  the  cavity scale must be  some multiple of the
wavelength.  

Geometrical models also face several obstacles before they can be
considered successful. The basic geometry is a challenge:    what
long-lived plasma structures can lead to the necessary 
interference or wave trapping?  In addition, the proportional band
spacing must be explained, perhaps by a variable index of
refraction in the interference or trapping region.

Geometrical models also need an underlying broad-band
radiation source, with at least 5 GHz bandwidth, in order to produce 
the emission bands we observe.  Because standard
pulsar radio emission mechanisms lead to relatively narrow-band
radiation, at the local plasma frequency, they seem unlikely to work
here.  A double layer might be the radiation source; charges accelerated
within the layer should radiate broadband, up to $\nu
\sim L / 2 \pi c$, if $L$ is the thickness of the acceleration region
within the double layer. Once again this is a small-scale effect; emission
at  10 GHz requires $L \sim 1$ cm.

\section{Final thoughts}

Our high time resolution observations of giant pulses from the Crab
pulsar have raised as many questions as they have answered.  The time
and frequency signatures of giant main pulses are consistent with
predictions of one current model of pulsar radio emission, namely,
strong plasma turbulence.  However, the time and frequency signatures
of giant interpulses  are totally different, and do not seem to match
the predictions of any current model.  This result is especially
surprising because magnetospheric models generally ascribe the main
pulse and the interpulse to physically similar regions, which simply
happen to be on opposite sides of the star.    One important clue may
be the offset in rotation phase between the high-radio-frequency interpulse,
and the interpulse which is seen at low radio frequencies and also in
optical and X-ray bands.  Does the high-frequency interpulse originate
in an unexpected part of the star's magnetosphere, where different
physical conditions produce such different radiation signatures?

\vskip 0.4cm

\begin{acknowledgements}
We appreciate helpful conversations with Joe Borovsky, Alice Harding,
Axel Jessner, 
Jan Kuijpers, Maxim Lyutikov, and the members of the Socorro pulsar
group.  This work was partially supported by the National Science
Foundation, through grant AST0139641 and through a cooperative
agreement with Cornell to operate the Arecibo Observatory.
\end{acknowledgements}
   


              \clearpage

\end{document}